# Guiding Quantitative MRI Reconstruction with Phase-wise Uncertainty

Haozhong Sun[1], Zhongsen Li[1], Chenlin Du[1],

Haokun Li[1], Yajie Wang[1], Huijun Chen[1]

[1] School of Biomedical Engineering, Tsinghua University
chenhj_cbir@mail.tsinghua.edu.cn

**Abstract.** Quantitative magnetic resonance imaging (qMRI) requires multi-phase acquisition, often relying on reduced data sampling and reconstruction algorithms to accelerate scans, which inherently poses an ill-posed inverse problem. While many studies focus on measuring uncertainty during this process, few explore how to leverage it to enhance reconstruction performance. In this paper, we introduce PUQ, a novel approach that pioneers the use of uncertainty information for qMRI reconstruction. PUQ employs a two-stage reconstruction and parameter fitting framework, where phase-wise uncertainty is estimated during reconstruction and utilized in the fitting stage. This design allows uncertainty to reflect the reliability of different phases and guide information integration during parameter fitting. We evaluated PUQ on in vivo T1 and T2 mapping datasets from healthy subjects. Compared to existing qMRI reconstruction methods, PUQ achieved the state-of-the-art performance in parameter mappings, demonstrating the effectiveness of uncertainty guidance. Our code is available at https://anonymous.4open.science/r/PUQ-75B2/.

**Keywords:** qMRI reconstruction • Uncertainty guiding • Parameter fitting • Deep learning.

## 1 Introduction

Quantitative magnetic resonance imaging (qMRI) quantifies tissue properties such as T1 and T2 relaxation times, enabling consistent measurements across different scanners and protocols. The primary approach to obtaining quantitative values involves acquiring multi-phase images to estimate parameters within a signal model of interest. However, this process requires multiple acquisitions, resulting in longer scan times compared to conventional MRI.

With the widespread adoption of deep learning (DL), numerous DL-based methods have been explored to accelerate qMRI. Most approaches [9,10,12,13] utilize convolutional neural networks (CNNs) to recover parameter maps from undersampled data. However, due to the ill-posed nature of this inverse problem and the complexity of CNN models, these acceleration methods introduce considerable uncertainty in their outputs. This uncertainty is particularly problematic for qMRI, as it directly affects tissue property measurements, which hold critical diagnostic value. While extensive research [1,8,18,24] has addressed uncertainty quantification in DL-based conventional

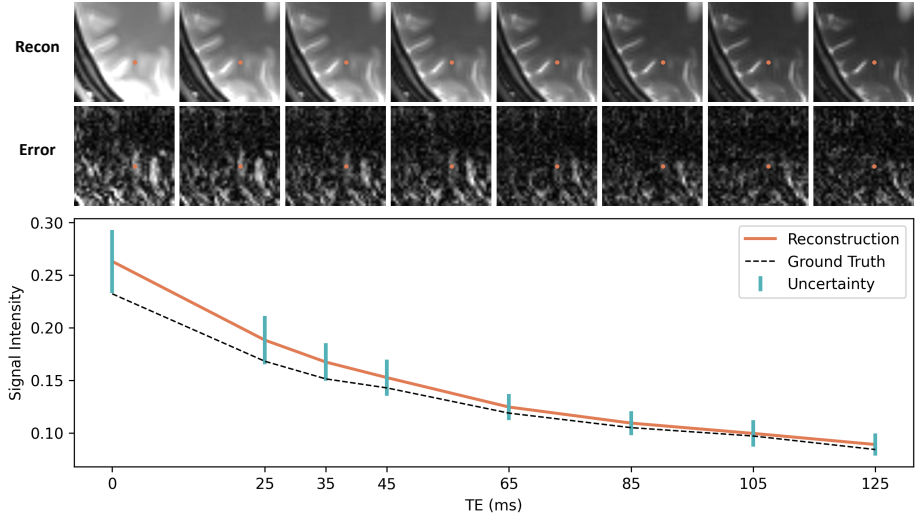

**Fig. 1.** Illustration of phase-wise uncertainty. The first row presents reconstructed images across eight TEs for T2 mapping, while the second row shows the corresponding residual errors. The bottom panel displays signal curves at the location marked by the orange point. The measured uncertainty reflects the reconstruction reliability across different phases.

MRI acceleration, studies on uncertainty in qMRI have also emerged [5,6,20]. However, existing work primarily focuses on quantifying uncertainty in the reconstruction process rather than leveraging it to improve reconstruction accuracy.

The challenge of utilizing uncertainty rather than measurement alone partly due to the nature of most uncertainty quantification approaches, which are simultaneous [4,7] with the prediction. In other words, by the time uncertainty is acquired, the reconstruction is already complete, which is why its application is often limited to downstream tasks [2,3,15,17]. However, qMRI inherently involves two tasks: undersampled data recovery and parameter estimation. This two-step reconstruction framework naturally provides an opportunity to integrate uncertainty. More importantly, aliasing from undersampling is a key source of uncertainty in qMRI. In practice, different contrast phases in qMRI exhibit distinct undersampling patterns for additional information [19], resulting in varying aliasing distributions, which also impact the reconstruction reliability of CNN models. Thus, the uncertainty from the phase dimension is particularly informative, as it can indicate which phases are more reliable for parameter estimation, as illustrated in Figure 1.

Based on the above insights, we propose a **P**hase-wise **U**ncertainty Guided **q**MRI reconstruction (PUQ) method, which employs a two-stage framework for reconstruction and parameter fitting, measuring uncertainty in different phases and using it to guide the integration of information during fitting. The proposed PUQ method was evaluated on an in-vivo dataset for both T1 and T2 mapping, where phase-wise uncertainty demonstrated its potential to enhance fitting accuracy. Across various undersampling factors, PUQ consistently achieved the lowest reconstruction error in parameter mapping compared to three existing qMRI acceleration methods [9,10,12]. To the best of our knowledge, this is the first approach to leverage uncertainty for improving MRI reconstruction accuracy.

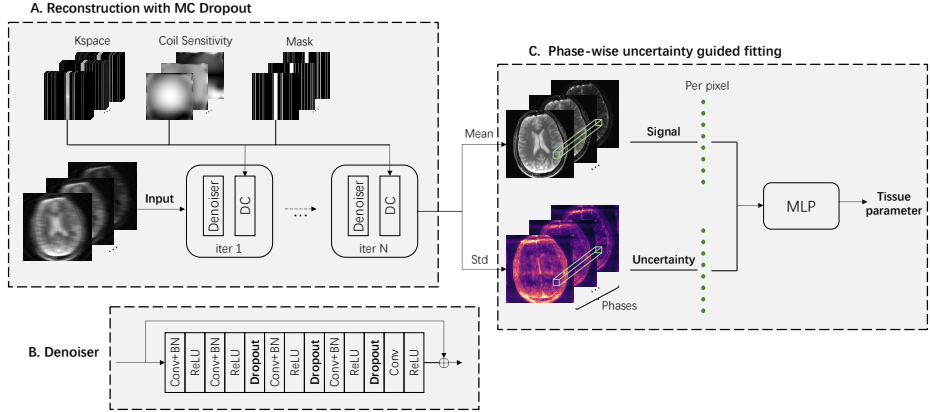

**Fig. 2.** The overall framework of proposed PUQ. A. Unrolled reconstruction model for multi-phase image reconstruction with MC Dropout. B. CNN Denoiser used in unrolled reconstruction model. C. Pixel by pixel parameter fitting guided by phase-wise uncertainty.

## 2 Methods

An overview of PUQ is shown in Fig. 2. The two-stage qMRI reconstruction consists of multi-phase image reconstruction and the uncertainty guided parameter fitting. An unrolled CNN with Monte Carlo (MC) dropout [4] is used to measure uncertainty for each phase in the recovered images. This uncertainty reflects the reliability of each phase, aiding in the effective information integration of signals across phases. Guided by the phase-wise uncertainty, parameter fitting is then performed pixel by pixel using a Multi-Layer Perceptron (MLP).

### 2.1 Reconstruction with MC Dropout

Figure 2(A) illustrates the unrolled model designed for multi-phase image reconstruction. The zero-filled images serve as inputs for five iterations (N=5) within a CNN denoiser and a data consistency (DC) layer, which projects the measured k-space data into the denoised images, following [21]. The denoiser weights are not shared across iterations. To facilitate Monte Carlo sampling, dropout layers are incorporated into the middle three hidden layers of the CNN denoiser, as depicted in Fig. 2(B). The image reconstruction problem could be seen as an approximation to such maximum a posteriori (MAP) inference for the predictive distribution:

$$p(x|x_u) = \int p(x|x_u,\theta)p(\theta|\mathcal{D})\,d\theta \tag{1}$$

where the x and $x_u$ represent the image and the paired zero-filled image, and $\mathcal{D}$ denotes the dataset. The posterior distribution $p(\theta|\mathcal{D})$ could be approximated by sampling network weight sample $\theta$ by dropped activations according to a Bernoulli distribution. With T samples drawn $(\theta_1, \ldots, \theta_T)$, the predictive mean is considered as the estimated images $\hat{x}_\mu$ and the uncertainty is represented by the standard deviation $\sigma$:

$$\hat{x}_\mu = \frac{1}{T}\sum_{t=1}^{T} f^{\theta_t}(x_u), \qquad \sigma = \sqrt{\frac{1}{T}\sum_{t=1}^{T}(\hat{x}_\mu - f^{\theta_t}(x_u))^2} \tag{2}$$

where $f$ denotes the forward pass of the unrolled model.

The hidden channel for denoiser is 64. The real and imaginary part of complex values are stacked along the channel dimension in the denoiser. The network is trained for 2000 epochs with dropout rate of 0.3 using the Adam optimizer [11], a batch size of 32, a learning rate of 0.01, and Mean Squared Error (MSE) as the loss function. Gradient Clip is used for avoiding exploding gradient with threshold of 0.01. During inference, dropout layers in the denoiser remain active for sampling. We perform 100 Monte Carlo samples to generate the mean reconstruction and the standard deviation as an uncertainty measure.

### 2.2  Phase-wise uncertainty guided fitting

To better utilize the uncertainty information from the reconstruction stage, parameter fitting in PUQ is performed pixel by pixel. For each pixel, the available inputs include the signal along the phase dimension and the corresponding uncertainty. A MLP with five fully connected layers and ReLU activations is used to fit the signal to the expected tissue parameter (T1 or T2 relaxation time in our experiments).

The MLP has 64 hidden units per layer and is trained for 200 epochs using the Adam optimizer [11] with a learning rate of 0.001 and a batch size of 1024, without using any regularization or data augmentation techniques. Mean squared error (MSE) loss is used for optimization. During both training and inference, the signal and uncertainty values are normalized by the signal value at the first phase.

## 3 Experiments

### 3.1 Dataset

To evaluate the proposed method, T2 and T1 mapping data were acquired from 20 healthy human subjects using a 3.0 T scanner (Ingenia CX, Philips Healthcare) with a 32-channel head coil. For T2 mapping, a 2D Turbo Field Echo (TFE) sequence with multiple T2 preparation pulses was used, where the T2 preparation time is referred to as TEprep. The imaging parameters were: TR = 2.8 ms, TE = 1.42 ms, TEpreps = [0, 25, 35, 45, 65, 85, 105, 125] ms, FA = 35°, bandwidth = 1085 Hz/pixel, field of view (FOV) = 200 × 200 mm², image matrix = 160 × 160, slice thickness = 8 mm, and 20 slices. For T1 mapping, a MOLLI [16] sequence with eight inversion time (TI) points was used. The scan parameters were: TR = 2.8 ms, TE = 1.39 ms, TIs = [251, 400, 1251, 1400, 2251, 2400, 3251, 4251] ms, FA = 20°, bandwidth = 1085 Hz/pixel, FOV = 200 × 200 mm², image matrix = 160 × 160, slice thickness = 8 mm, and 20 slices.

Table 1. Quantitative results on T2 and T1 datasets under different acceleration rates.

|   | Methods | 6× | | 8× | | 10× | |
|---|---|---|---|---|---|---|---|
|   |   | NRMSE | SSIM | NRMSE | SSIM | NRMSE | SSIM |
| T2 | **MANTIS** [12] | 0.4025 | 0.9293 | 0.4198 | 0.9255 | 0.4292 | 0.9203 |
|   | **Dopamine** [10] | 0.3944 | 0.9364 | 0.4181 | 0.9273 | 0.4253 | 0.9259 |
|   | **DeepT1** [9] | 0.3173 | 0.9584 | 0.3476 | 0.9493 | 0.3589 | 0.9448 |
|   | **PUQ (ours)** | **0.2798** | **0.9694** | **0.3128** | **0.9607** | **0.3291** | **0.9563** |
|   | **w/o G** | 0.2874 | 0.9681 | 0.3151 | 0.9605 | 0.3348 | 0.9554 |
| T1 | **MANTIS** [12] | 0.1333 | 0.9472 | 0.1609 | 0.9258 | 0.1768 | 0.9148 |
|   | **Dopamine** [10] | 0.1642 | 0.9322 | 0.1699 | 0.9224 | 0.1854 | 0.9084 |
|   | **DeepT1** [9] | 0.0588 | 0.9839 | 0.0844 | 0.9701 | 0.1047 | 0.9566 |
|   | **PUQ (ours)** | **0.0475** | **0.9893** | **0.0739** | **0.9769** | **0.0932** | **0.9650** |
|   | **w/o G** | 0.0481 | 0.9891 | 0.0748 | 0.9764 | 0.0938 | 0.9645 |

In total, the T2 and T1 datasets contain 400 image slices, split into 280/60/60 for training, validation, and testing. All images were compressed to eight coils, with coil sensitivity estimated using ESPIRiT [22]. The ground truth parameter mappings were obtained using direct least squares fitting of signal evolution. The images were then retrospectively undersampled using a 1D Cartesian random pattern with varying acceleration factors.

### 3.2 Evaluation

We compared the proposed PUQ with three existing qMRI reconstruction methods [9,10,12]: MANTS [12] and Dopamine [10], which use a one-step approach for reconstruction and parameter fitting, and DeepT1 [9], which follows a two-step framework similar to PUQ. To quantitatively assess the effectiveness of parameter map reconstruction, we compared the normalized root mean square error (NRMSE) and structural similarity (SSIM) values. All reconstruction methods were implemented in PyTorch on an Ubuntu 20.04 LTS system with eight NVIDIA A800 GPUs (80 GB each).

### 3.3  Results

First, we compared PUQ with the other methods and a version of PUQ without uncertainty guidance (denoted as w/o G) on the T1 and T2 datasets. A 1D Cartesian undersampling pattern was used, with acceleration rates of [6, 8, 10]. The Auto-Calibration Signal (ACS) area was set to 0.06 for acceleration rates of 6 and 8, and 0.08 for 10.

The quantitative results are presented in Table 1. In both T1 and T2 datasets, PUQ achieved the best NRMSE and SSIM across all acceleration rates compared to the three other methods. Compared to its w/o G variant, PUQ consistently improved SSIM while slightly increasing NRMSE across all acceleration rates. We also observed that the two-stage method, DeepT1, outperformed the one-stage methods, MANTS and Dopamine, likely due to the reduced complexity offered by the two-stage design. The qualitative results are shown in Fig. 3, where, due to space constraints, only the 6× acceleration example is displayed. For both T1 and T2 datasets, PUQ estimated parameter maps with the lowest residual error. In the T2 example, PUQ effectively distinguished white and gray matter, while in the T1 example, it more accurately recovered tissue border regions with varying T1 values. Across different acceleration rates, the proposed PUQ achieved the highest performance among existing methods, with uncertainty guidance showing effectiveness for T2 and T1 mapping.

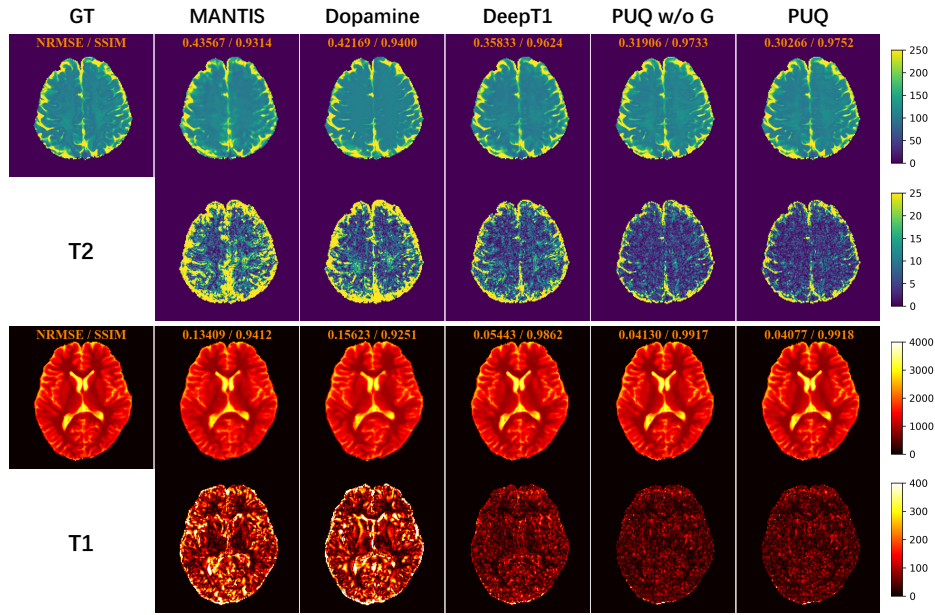

**Fig. 3.** Visual comparison with color bars and error maps on the T2 and T1 dataset. The NRMSE and SSIM of the example are shown in the top of reconstructed parameter mappings.

In PUQ, uncertainty in the reconstruction stage is measured using MC Dropout, which inherently influences reconstruction performance. We thus examined the effects of dropout rate and sampling times as hyperparameters in MC Dropout. The dropout rate experiment was conducted on the T1 dataset at 8× acceleration rate, while the sampling times experiment was performed on the T2 dataset with a 10× acceleration rate.

The results are shown in Fig. 4. As seen in Fig. 4A, we explored sampling times of [10, 20, 50, 100, 200]. Increasing the number of samples generally reduced NRMSE for methods without uncertainty guidance, except for 200 samples, which slightly increased NRMSE compared to 100. This may be due to averaging sufficiently eliminating random errors. Notably, across all sampling times, PUQ (w/ G) consistently achieved lower errors than its unguided counterpart, demonstrating the effectiveness of MC Dropout-based uncertainty measurement. Fig. 4B presents results for dropout rates of [0.2, 0.3, 0.4]. Higher dropout rates led to decreased performance, with a dropout rate of 0.4 resulting in higher NRMSE than the model without dropout. However, uncertainty guidance remained effective across different dropout rates, consistently improving performance.

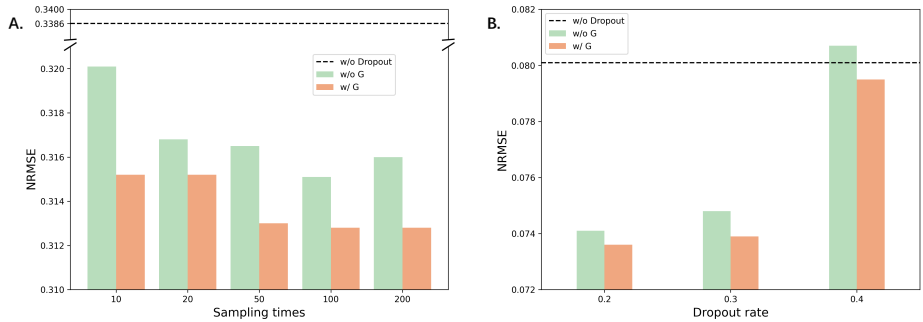

**Fig. 4.** Results with different hyper-parameter of MC Dropout, containing sampling times at A and Dropout rate at B.

### 3.4  Ablation Study.

To assess the effectiveness of phase-wise uncertainty guidance, we compared the model without uncertainty guidance ("w/o G") and the model without MC Dropout in the reconstruction stage ("w/o Dropout"). This ablation study was conducted using a 1D Cartesian undersampling pattern at an 8× acceleration rate. Additionally, since MC Dropout relies on sampling approximation, model performance may be influenced by sampling randomness. To account for this, we repeated each model's training and testing five times and reported the mean performance and standard deviation in Table 2. While improving baseline performance was not our primary goal, MC Dropout itself enhanced the reconstruction quality, likely due to reduced overfitting. Notably, the proposed method with uncertainty guidance achieved the lowest NRMSE and highest SSIM on average. Moreover, it consistently outperformed the version without guidance in every repetition (not shown in Table), demonstrating the value of pixel-wise uncertainty for parameter mapping.

Furthermore, we explored an alternative uncertainty measurement approach based on negative log-likelihood (NLL) [23] minimization, which is optimized via gradient descent to estimate the observation noise parameter of a heteroscedastic Gaussian distribution, serving as a measure of uncertainty. We examined its combination with MC Dropout [7], resulting in NLL+MD w/o G (without uncertainty guidance) and NLL+MD w/ G (with uncertainty guidance). We then evaluated NLL separately, where NLL w/ G denotes the version using NLL-based uncertainty for guidance, and NLL w/o G refers to the version without guidance. The results, presented in the lower part of Table 2, indicate that the uncertainty guidance from the NLL approach is effective. Whether combined with MC Dropout or not, the guided versions achieved lower

Table 2. Ablation results on T2 and T1 datasets

| Methods | T2 | | T1 | |
|---|---|---|---|---|
| | NRMSE | SSIM | NRMSE | SSIM |
| w/o Dropout | 0.3304 ± 0.0643 | 0.9566 ± 0.0164 | 0.08138 ± 0.0117 | 0.9747 ± 0.0111 |
| w/o G | 0.3175 ± 0.0640 | 0.9599 ± 0.0153 | 0.07618 ± 0.0113 | 0.9756 ± 0.0109 |
| **Proposed** | **0.3125 ± 0.0623** | **0.9605 ± 0.0151** | **0.07569 ± 0.0115** | **0.9759 ± 0.0109** |
| NLL w/o G | 0.3302 ± 0.0657 | 0.9571 ± 0.0161 | 0.08401 ± 0.0116 | 0.9744 ± 0.0109 |
| NLL w/ G | 0.3279 ± 0.0643 | 0.9573 ± 0.0159 | 0.08365 ± 0.0115 | 0.9746 ± 0.0108 |
| NLL+MD w/o G | 0.3306 ± 0.0627 | 0.9568 ± 0.0155 | 0.07886 ± 0.0109 | 0.9759 ± 0.0103 |
| NLL+MD w/ G | 0.3240 ± 0.0597 | 0.9577 ± 0.0150 | 0.07759 ± 0.0109 | 0.9767 ± 0.0100 |

NRMSE. It is widely accepted that the MC Dropout reflects epistemic uncertainty, while NLL-based uncertainty primarily captures aleatoric uncertainty. This experiment suggests that both types of loss encode valuable pixel-wise reliability information for improved fitting. However, integrating NLL loss itself significantly degraded performance, likely due to its strong dependence on the predictive variance gradients [19]. This degradation outweighed the benefits of uncertainty guidance, making it unsuitable for qMRI reconstruction.

## 4    Conclusion

In this work, we introduced the uncertainty-guided model PUQ for qMRI reconstruction. PUQ proposes a two-stage framework that uniquely leverages uncertainty information to enhance parameter mapping performance. In comparisons with other qMRI reconstruction methods, PUQ achieved the best reconstruction metrics and enhanced the quality of estimated parameter mappings in T1 and T2 datasets.

The uncertainty in PUQ is measured during the initial step of multi-phase image reconstruction. We use this uncertainty to better incorporate information from different phase channels for parameter estimation. Comparisons at various acceleration rates demonstrated the effectiveness of uncertainty guiding. Additionally, dropout, a widely used method in neural networks, also impacts performance. Our experiments show that across different dropout rates and sampling times, PUQ consistently provides stable performance improvements. The ablation study with repeated training and testing further underscores the effectiveness of uncertainty for qMRI reconstruction.

We believe that the proposed PUQ has significant potential for broader applications that could utilize uncertainty information, such as MR fingerprinting, where incorporation of data from multiple phases is also required. The uncertainty from different phases reflects the reliability of each phase, facilitating better information fusion.